\documentclass[conference]{IEEEtran}

\usepackage[toc]{glossaries}

\usepackage{ifpdf}
\usepackage{graphicx}
\usepackage{float}
\usepackage{dsfont}
\usepackage{xcolor}
\usepackage{comment}
\usepackage{mathtools, cuted}
\usepackage{ctable}
\usepackage{multicol}
\usepackage[ruled]{algorithm2e}
\usepackage[margin=0cm]{caption}
\usepackage{subcaption}
\usepackage{soul}
\usepackage[normalem]{ulem}
\usepackage{multirow}
\usepackage{tablefootnote}
\usepackage{enumitem}

\usepackage{cite}

\usepackage{amsmath}
\usepackage{amssymb}
\usepackage{amsthm}
\usepackage{mathdots}
\usepackage{amsfonts}
\usepackage[long]{optidef}

\usepackage{algorithmic}

\usepackage{array}

\makeatletter \def\algocf@capseparator{}\makeatother

\usepackage{stfloats}

\def\BibTeX{{\rm B\kern-.05em{\sc i\kern-.025em b}\kern-.08em
    T\kern-.1667em\lower.7ex\hbox{E}\kern-.125emX}}

\begin{document}

\title{The impact of mobility, beam sweeping and smart jammers on security vulnerabilities of 5G cells}


 \author{\IEEEauthorblockN{Ghazal Asemian, Michel Kulhandjian, Mohammadreza Amini,\\ Burak Kantarci, Claude D'Amours and Melike Erol-Kantarci}\vspace{-3mm}\\
 \IEEEauthorblockA{School of Electrical Engineering and Computer Science, University of Ottawa, Ottawa, ON, Canada \\
 \texttt{\{gasem093,mkulhand,mamini6,burak.kantarci,cdamours,melike.erolkantarci\}@uottawa.ca}}
 \vspace{-0.25in}
 }

\maketitle

\begin{abstract}

The vulnerability of 5G networks to jamming attacks has emerged as a significant concern. This paper contributes in two primary aspects. Firstly, it investigates the effect of a multi-jammer on 5G cell metrics, specifically throughput and goodput. The investigation is conducted within the context of a mobility model for user equipment (UE), with a focus on scenarios involving connected vehicles (CVs) engaged in a mission. Secondly, the vulnerability of synchronization signal block (SSB) components is examined concerning jamming power and beam sweeping. Notably, the study reveals that increasing jamming power beyond $40$ dBm in our specific scenario configuration no longer decreases network throughput due to the re-transmission of packets through the hybrid automatic repeat request (HARQ) process. Furthermore, it is observed that under the same jamming power, the physical downlink shared channel (PDSCH) is more vulnerable than the primary synchronization signal (PSS) and secondary synchronization signal (SSS). However, a smart jammer can disrupt the cell search process by injecting less power and targeting PSS-SSS or physical broadcast channel (PBCH) data compared to a barrage jammer. On the other hand, beam sweeping proves effective in mitigating the impact of a smart jammer, reducing the error vector magnitude root mean square from $51.59\%$ to $23.36\%$ under the same jamming power.
\end{abstract}


\IEEEpeerreviewmaketitle

\section{Introduction}
\label{Introduction} 

The susceptibility of 5G NR to external attacks, particularly jamming, is a significant concern given its inherent nature of wireless radio frequency transmission \cite{cao2020,Sullivan2021}. Jammers have the potential to deplete substantial resources and disrupt critical applications, posing serious risks to areas like national defense, self-driving technology, public safety, and healthcare, as 5G and beyond infrastructures become increasingly essential for government agencies and commercial businesses \cite{Chorti2022}. The rise in electronic attacks on 5G networks, with jamming being a predominant method, has been noted \cite{cao2020}. 

Jamming attacks including proactive, reactive, barrage, function-specific, and protocol-specific \cite{Pirayesh2022, Mpitziopoulos2009} manifest at various levels, including physical, network, and application layers, ranging from radio frequency interference that blocks wireless transmission to the distortion of packets in legitimate communications. These attacks range from blocking wireless transmission through radio frequency interference to distorting packets in legitimate communications. Notably, the synchronization signal (SS) block is a crucial component of the 5G waveform, and attacks on this block can lead to denial of service, increased overhead in terms of re-transmission, and heightened power consumption \cite{birutis2022practical}. In comparison to barrage jammers, which apply jamming signals across the entire 5G resource grid, attackers can enhance the impact and efficiency of their attacks by specifically targeting the SS block.

To address these concerns, we investigate the characteristics of barrage jamming and protocol-specific jamming attacks, utilizing the spatio-temporal parametric stepping (STEPS) mobility model \cite{nguyen2011steps}. Our specific contributions are outlined as follows:

      \vspace{-0.0cm}
      \begin{enumerate}[label=(\arabic*)]
      \item Study the impact of multi-jammer attack scenarios on 5G cell metrics, including throughput and goodput.
       \item Examine the impact of mobile user equipment (UE), particularly utilizing the STEPS mobility model, along with the spatial placement of jammers to achieve maximum disruption of cell metrics..
       \item Investigate jamming attacks on the 5G synchronization signal block (SSB), various reference signals such as primary synchronization signal (PSS), secondary synchronization signal (SSS), and demodulating reference signal (DM-RS) in the physical broadcast channel (PBCH). This exploration encompasses the extraction of the physical cell identity (PCI) and master information block (MIB).
      \end{enumerate}
 \vspace{-0.0cm}

The paper is organized as follows: Section \ref{RelatedWork} introduces related works, while Section \ref{sec_system} discusses the network, mobility, channel, and jammer models. Section \ref{sec_performance} presents the performance analysis of the network under jamming, including the vulnerability of the SSB concerning jamming power. Finally, Section \ref{sec_conclusion} provides the conclusions.

\section{Related Work}
\label{RelatedWork}
Several studies have explored jamming attacks in 5G networks, categorizing them based on their strategies to disrupt legitimate signals \cite{arjoune2020smart,lichtman20185g,flores2023implementation,grover2014jamming,Basciftci2015,Angueira2022}. While individual jamming attacks have been extensively studied, there is a need to investigate the impact of multiple simultaneous jamming attacks on 5G networks \cite{Thai2022}. Understanding this scenario is crucial for developing more robust detection and localization algorithms tailored to critical 5G scenarios. Simultaneous attacks can lead to severe consequences, with longer jamming durations and reduced resource usage, posing challenges for detection and localization techniques.

Tague \textit{et al.} \cite{tague2008linear} investigated flow-jamming attacks in simple wireless networks, focusing on metrics such as jamming impact, efficiency, and resource variation using linear programming. In a flow-jamming attack, adversaries strategically place multiple jammers within a network to disrupt the flow of communication. This type of attack is categorized into two forms based on their attack control entity: a centralized flow-jamming attack, where a central control entity is involved, and a distributed flow-jamming attack, where jammers operate autonomously without centralized control. Cheng \textit{et al.} \cite{cheng2014m} investigated the impact of multi-jammer attacks in wireless sensor networks (WSNs) and proposed two localization algorithms: M-cluster and X-ray. In their proposed topology, sensor nodes in the network are categorized as jammed, unaffected, or boundary nodes. The increasing number of jammers leads to broader jammed regions, amplifying the jamming effect and posing challenges for network defenders. While studies like \cite{tague2008linear} and \cite{cheng2014m} establish the foundation to understand the impacts of jamming in simpler network scenarios, and proposed preliminary models for jammer localization, recent developments focus on the complex dynamic 5G network configuration. For instance, Liu \textit{et al.} \cite{liu2013error} introduced a localization method to locate multiple jammers in wireless networks by computing jamming signal strength (JSS) using ambient noise floor (ANF). Also, Atya \textit{et al.} \cite{atya2015exploiting} implement a jamming mitigation technique for wireless networks based on the 802.11a standard. This technique aims to preserve the network's throughput even in the presence of a jamming attack. The paper evaluates the performance of the mitigation technique in the presence of multiple jammers in a network with 50 nodes.
 
In addition to the number of jammers in a jamming attack scenario, the impact of the attack can be amplified by targeting vulnerable and critical control signals in the 5G resource grid.
Some works that investigate the vulnerability of critical 5G NR channels include \cite{Wang2023, arjoune2020smart, lichtman20185g}. Wang \textit{et al.} \cite{Wang2023} propose a scheme to detect an intelligent jammer attacking the PBCH using principal component analysis. This attack disrupts the MIB recovery and cell connection process. The proposed method's advantage is its independence from any attacker information, as it employs adaptive thresholding computed from statistics. The vulnerability of PBCH and the physical downlink control channel (PDCCH) in 5G new radio (NR) to selective jamming attacks is discussed in \cite{arjoune2020smart}. This design flaw becomes particularly alarming when higher-frequency carriers are considered, requiring the jammer to be in close proximity to the mobile station for an effective attack. Lichtman \textit{et al.} \cite{lichtman20185g} emphasize the vulnerability of 5G NR channels, including PDCCH, PBCH, and physical downlink shared channel (PDSCH), to jamming attacks. They introduce PBCH jamming and PSS jamming as potential high-impact attacks, underscoring the need for specific considerations to enhance the security of 5G NR. This study investigates vulnerabilities arising from the connection between implementations of the 5G NR and long term evolution (LTE) protocols.
These works emphasize the need for specialized countermeasures to protect against both broad and targeted jamming attacks.

Although localization techniques can identify the jammer's location, mobile jammers remain a significant threat, particularly in vehicular networks and unmanned aerial vehicular networks, where rapid transmission restoration is crucial \cite{Pirayesh2022, Karagiannis2018}. Consequently, localizing the jammer is imperative for implementing security measures against the jammer and restoring transmission in such mobile environments. Rani Dey \textit{et al.} \cite{Meenu2023}, propose a real-time mechanism for detecting and localizing denial of service (DoS) attacks in vehicular networks. Their approach utilizes data packet counters and average packet delivery ratio (PDR), augmented with a supervised machine learning-based solution to enhance robustness and consistency. By leveraging PDR and triangulation-based methods, they successfully localize both intentional and unintentional DoS attacks.
 
The interaction between mobility and jamming is explored in \cite{Balakrishnan2019, Malebary2016}, where the authors examine the influence of movement patterns on the effectiveness of jamming and the strategies for its mitigation.
Balakrishnan \textit{et al.} \cite{Balakrishnan2019}, analyze the impact of mobility on physical layer security using an analytical model for secrecy metrics, focusing specifically on mmWave users under the random waypoint mobility model. Meanwhile, Malebary \textit{et al.} \cite{Malebary2016}, investigate the effects of jamming attacks under mobility and behavior in IEEE802.11p networks. Their study evaluates jamming effectiveness under different mobility patterns and proposes a jammer detection scheme tailored for IEEE802.11p networks. Additionally, \textit{et al.} \cite{punal2015}, conduct an extensive investigation into interference mitigation techniques, assessing the impact of reaction delay and interference signal length on car-to-car communications. These studies provide valuable insights into mitigating interference and enhancing security in vehicular networks. We summarize and compare our contribution with existing studies in Table \ref{GapAnalysis}. 
 \begin{table*}
 \vspace{.032in}
 \sffamily
 	\renewcommand{\arraystretch}{1.6}
 \caption{Gap analysis and contributions of this work with respect to the existing studies in the literature.} 
\begin{center}
    \begin{tabular}{ | l | c | c | c | c | c | c | c | c | c | c | }
    \hline
    \label{GapAnalysis}
    \textbf{Contributions} & \textbf{This work} &  \cite{Bouzabia2023} & \cite{Dinh2023} & \cite{Venkata2021} & \cite{Wang2023} & \cite{Karagiannis2018} & \cite{Cheng2012} & \cite{Balakrishnan2019} & \cite{Malebary2016} & \cite{punal2015}\\ \hline \hline
    Detection Algorithm & \checkmark & \checkmark & \checkmark & \checkmark  & \checkmark &  \checkmark & \checkmark &  & \checkmark &  \\ 
     \hline
     Mobility  & \checkmark &  & \checkmark & \checkmark & &  \checkmark &  & \checkmark & \checkmark & \checkmark \\ \hline
    Presence of Multi-Jammers & \checkmark &  &  &  &  & & & & &   \\ \hline
    Attacks on Different Reference Signals & \checkmark &  &   &  &  &  &  &  &  & \\ \hline
    \end{tabular}
    \end{center}
    \vspace{-6mm}
    \end{table*}

This study delves into the intricate aspects of 5G networks when subjected to jamming attacks. It incorporates a mobility model to illustrate the movement of UEs across the network and areas affected by jamming. Signal attenuation is considered in the jammer models to simulate the impact of jammer locations in the network. Additionally, the study accounts for multi-jammer scenarios, discussing the positions of jammers, their overlapping areas, and the effective number of jammers. This investigation contributes to a deeper understanding of the susceptibility of 5G networks to jamming attacks and provides insights that can inform the development of robust jamming detection and mitigation techniques.

\section{System Model}
\label{sec_system}
\subsection{5G Network Model}

We employ multiple UEs and multiple gNodeBs (gNBs) to simulate multi-user scenarios in 5G networks. To comprehensively assess the impacts of inter-cell interference and user mobility, our study establishes a dynamic network configuration that allows UEs to move freely within their designated cell areas, introducing a layer of complexity reflective of real-world scenarios. Each gNB is configured to cover a distance of $R_{cell}$ with a transmit power of $P_{gNB}^{Tx}$. Within the coverage area of each cell, $n_{UE}$ UEs are randomly positioned to capture the heterogeneity and distribution of users in 5G networks. UEs are equipped with a single antenna for communication, enabling them to receive digitally modulated transmitted signals.

In addition, $n_{RB}$ physical resource blocks are considered for data transmission and reception. Each resource block comprises 12 subcarriers with a spacing of $\Delta f$ for frequency separation between neighboring subcarriers. Furthermore, $n_{RB}^{ssb}$ resource blocks, as per the standard, are assigned to the SSB. Parameters such as downlink carrier frequency $F_c^{DL}$, downlink bandwidth $BW_{DL}$, uplink carrier frequency $F_c^{UL}$, and uplink bandwidth $BW_{UL}$ define the central frequency and spectral space for both downlink and uplink transmissions. The downlink application data rate $C_{app}^{DL}$ quantifies the total amount of data reliably transmitted from the gNB to the UEs within a given timeframe.

\subsection{Mobility Model}
To address the challenges associated with traditional mobility models like random waypoint and random walk in the context of user mobility, this study adopts the STEPS model \cite{nguyen2011steps}. The STEPS model offers a versatile framework for simulating diverse human mobility patterns, allowing manipulation through a concise set of parameters. Incorporating the principles of preferential attachment and location attraction, the STEPS model captures intrinsic spatio-temporal correlations in human mobility behaviors. Notably, the model leverages a power law distribution to govern both spatial attraction and temporal preference, enhancing adaptability. The probability density function (PDF) of this power law distribution is defined as \vspace{-2mm}
\begin{equation}\label{eq:powerLaw}
    \begin{split}
      P[D=d] = \frac{\beta}{{(1+d)}^\alpha}
    \end{split} ,
\end{equation}
\noindent where $d$ represents the distance from the preferred zone, $\beta$ is a normalizing constant ensuring the PDF integrates to 1 at all distances, and $\alpha$ is the power law exponent determining the intensity of zone attraction. The power law also dictates the duration a node stays in its preferred zone, expressed as \vspace{-2mm}
\begin{equation}\label{eq:powerLawTime}
    P[T=t] = \frac{\omega}{t^\tau} ,
\end{equation}
\noindent where $\tau$ and $\omega$ are parameters that model the staying time $T$ of a node in a particular zone within the mobility model. $\tau$ denotes the temporal preference level of the node, reflecting the degree to which the node prefers to remain within a specific zone and $\omega$ is the normalization factor ensuring the PDF integrates to a total of 1. Adjusting the power law factor allows for the generation of diverse mobility patterns, ranging from completely random to highly localized.

\subsection{Channel Model}
This study explores multipath propagation in urban environments by employing the widely adopted clustered delay line (CDL) model. The CDL model amalgamates paths with distinct delay spread ($DS$) and angle-of-arrival characteristics, effectively capturing complex signal interactions within urban settings, encompassing both line-of-sight (LOS) and non-line-of-sight (NLOS) components.
The received signal, traversing the CDL channel, accounts for signal intensity variation caused by clusters' impairments in the environment. Additionally, the channel model includes the fundamental free space path loss (FSPL), which factors in the decrease in signal intensity due to distance. Based on the FSPL model, the received signal power is expressed as follows,
\vspace{-2mm}
\begin{equation}\label{eq:RxWaveform}
    P_r = P_t + G_t + G_r - 20 \log{[\frac{ {\lambda}^2}{{4\pi}^2d^2}]} ,
\end{equation}
\noindent where $P_t$ represents the transmitted power, $G_t$ and $G_r$ denote transmitter and receiver antenna gains, and $\lambda$ and $d$ represent wavelength and transmitter-receiver distance, respectively. The incorporation of the CDL model and FSPL contributes to the understanding of interference and channel attenuation dynamics.

\subsection{Jammer Model}
A barrage jammer emits Gaussian noise across the entire 5G downlink bandwidth. The FSPL model is employed to consider signal attenuation and distance-related propagation effects. To simulate a real-world jamming attack more accurately, the jamming signal undergoes transmission through the CDL channel model, accounting for the time-varying nature of wireless communication and incorporating precise signal distortions. Similar to the received signal power in (\ref{eq:RxWaveform}), the power received by the user equipment (UE) from the jammer ($P_j^{rx}$) is expressed as
\begin{equation}\label{eq:JSS}
    P_j^{rx} = P_j +G_j +G_r- 20 \log{[\frac{ {\lambda}^2}{{4\pi}^2{d_j}^2}]} ,
\end{equation}
\noindent where $P_j$ denotes the jammer's transmit power, and $G_j$ and $d_j$ represent the gain of the jammer antenna and the distance of a node to the jammer, respectively.

Moreover, a smart jammer in this context is defined as a jammer that strategically transmits its power on a specific portion of the resource grid (RG), thereby making its actions more energy-efficient.

\section{Performance analysis of mobility, jammer position, and targeted reference
signals}
\label{sec_performance}
In this study, the MATLAB 5G Toolbox is employed to assess the impact of a jamming attack within the dynamic context of a multi-user 5G network. The evaluation takes into account factors such as UE mobility, jammer position, and targeted reference signals.
The simulation inputs are determined based on parameters outlined in 3rd generation partnership project (3GPP) specifications \cite{3GPPTS138104, 3GPPTS138901, 3GPPTS138101-1, 3GPPTS138323}, while the transmit power\footnote{This power is correspondence to effective isotropic radiated power (EIRP).} of the gNB is adapted from the findings in \cite{han2021hybrid}. The simulation is operated on the 5G NR n7 band with a center frequency of 2.635 GHz based on the 3GPP specifications. For clarity, Table \ref{table:parameters} provides a comprehensive list of parameters and assumptions utilized in the simulation scenario, as discussed in Section \ref{sec_system}.

\begin{table}[h!] 
\vspace{.032in}
\centering
\renewcommand{\arraystretch}{1.45}
\caption{Simulation settings }
\begin{tabular}{||m{6em}| m{1.15cm} || m{7em}| m{1.25cm} ||} 
 \hline
 Parameter & Value & Parameter & Value\\ [0.5ex] 
 \hline\hline
  $n_{UE}$ & 20 & $R_{cell}$ & 500 m  \\
   \hline
   $P_{gNB}^{Tx}$ &  32 dBm & $n_{RB}^{ssb}$ & $20$\\
   \hline
   $n_{RB}$ & 51 & $n_{RE}$\tablefootnote{$n_{RE}$ represents the number of resource elements.} & 12 $n_{RB}$ \\
   \hline
   $\Delta f$ & 30 kHz & $F_c^{DL}$ & 2.635 GHz\\
   \hline
   $BW_{DL}$ & 20 MHz & $C_{app}^{DL}$ & 16 kbps\\
   \hline
   Duplex Mode & FDD & $F_c^{UL}$ & 2.515 GHz\\
   \hline
   $BW_{UL}$ & 20 MHz & RLC SDU Size & 9 kb\\
   \hline
   Channel Model & CDL-A & DS & 30 ns\\
   [1.5ex] 
 \hline
\end{tabular}
\label{table:parameters}
\vspace{-5mm}
\end{table}

\subsection{Cell Network Metrics Under Jamming}

The movement of UE induces changes in path loss values and received signal strength (RSS), directly influencing the received signal quality. Signal strength variations, such as a decrease in signal strength, result in a diminished signal-to-noise ratio (SNR), thereby elevating the likelihood of errors and packet loss. In scenarios involving a jamming attack, where intentional interference disrupts the communication link, these fluctuations in signal intensity become critical.

As the signal strength of a UE diminishes due to path loss, its susceptibility to the effects of the jamming attack increases. Interference signals can further deteriorate the legitimate signal, making it challenging for the UE to accurately receive and decode data. The impact of the jamming attack can vary over time and space as the UE moves and encounters fluctuations in signal intensity. This dynamic nature underscores the importance of considering both spatial and temporal dimensions when assessing the effectiveness and consequences of jamming attacks on the communication system.

To comprehensively assess the impact of a jamming attack within a realistic multi-cell/multi-user UE 5G NR scenario, an end-to-end link-level simulation has been established. This simulation takes into account UE mobility and integrates all functions associated with the jamming scenario, spanning both the physical layer (L1) and the medium access control layer (MAC - L2). Key components, including the MAC scheduler, modulation and coding scheme (MCS), hybrid automatic repeat request (HARQ), and channel state information (CSI) configuration, are considered.

The evaluation of network metrics, cell throughput, and goodput provides insights into the performance of the communication system under the influence of a jamming attack. This holistic simulation framework enables a thorough examination of the interplay between various system functions and their collective impact on network performance in the presence of jamming-induced challenges.
. 

\begin{figure}[htp]
    \centering
    \includegraphics[width=6.5cm]{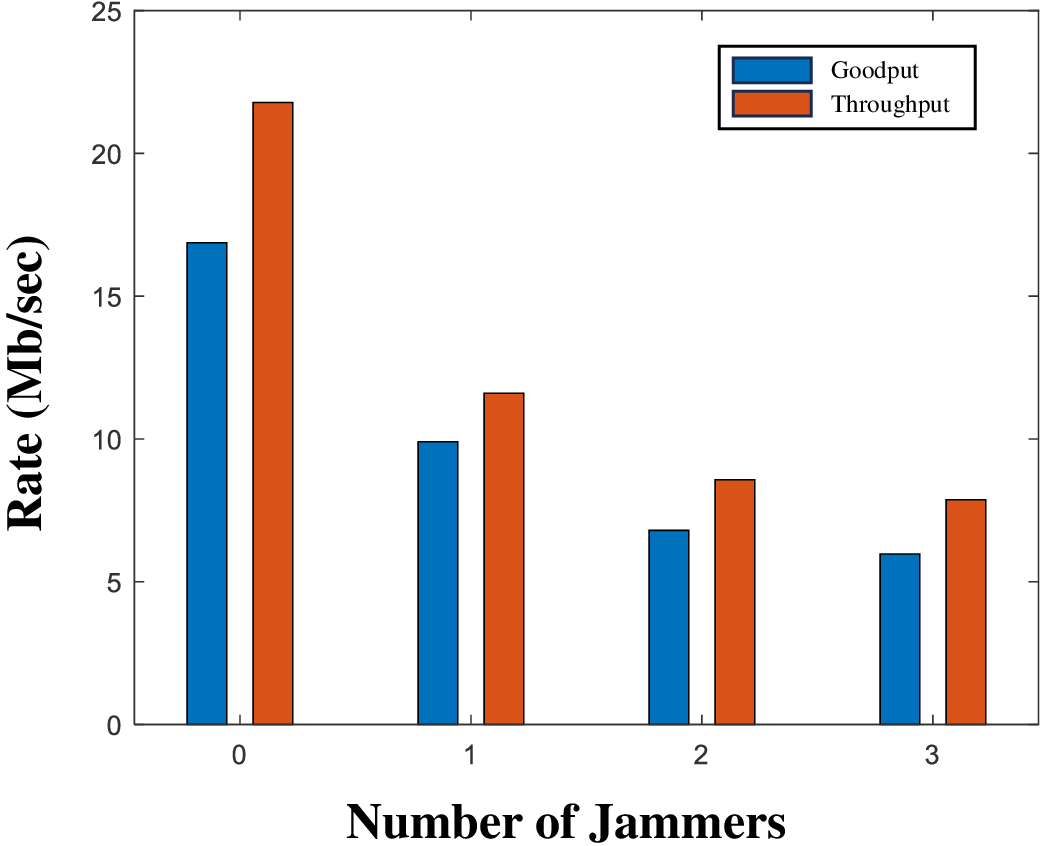}
    \caption{Cell throughput and goodput vs. number of jammers with jammer distance of $224 \: m$ at jamming power of $20$ dBm.}
    \label{fig:n of jammer}
    \vspace{-2mm}
\end{figure}

The impact of network metrics in relation to the number of jammers is visually represented in Fig. \ref{fig:n of jammer}. To specifically investigate the influence of varying numbers of jammers, the distance is fixed at $224$ meters. This approach ensures a focused examination of the effects of changing jammer quantities while minimizing distance-related factors' impact on the results. The goal is to avoid variations in jamming signal strength due to distance fluctuations.

With a single jammer present, both throughput and goodput experience a significant decrease, with throughput dropping by $46.78\%$. Introducing a second jammer further exacerbates the degradation of cell throughput, resulting in an additional $26.12\%$ decline. However, the introduction of a third jammer leads to a marginal deterioration in cell metrics, approximately $8.16\%$, indicating an overlapping affected area. In a realistic scenario involving multiple jammers within a 5G cell, the throughput degrades even with the presence of a single jammer, as verified in Fig. \ref{fig:n of jammer}, until reaching a certain threshold of jammer density. This saturation of jamming effects occurs because the network's inherent capacity and resilience mechanisms are severely compromised by the initial set of jammers. Operating near its interference threshold, the network has limited tolerance for additional jammers, diminishing the contribution room for further disruptions.

\graphicspath{{figs/}}

\begin{figure}[htp]
    \centering
    \includegraphics[width=7.95cm]{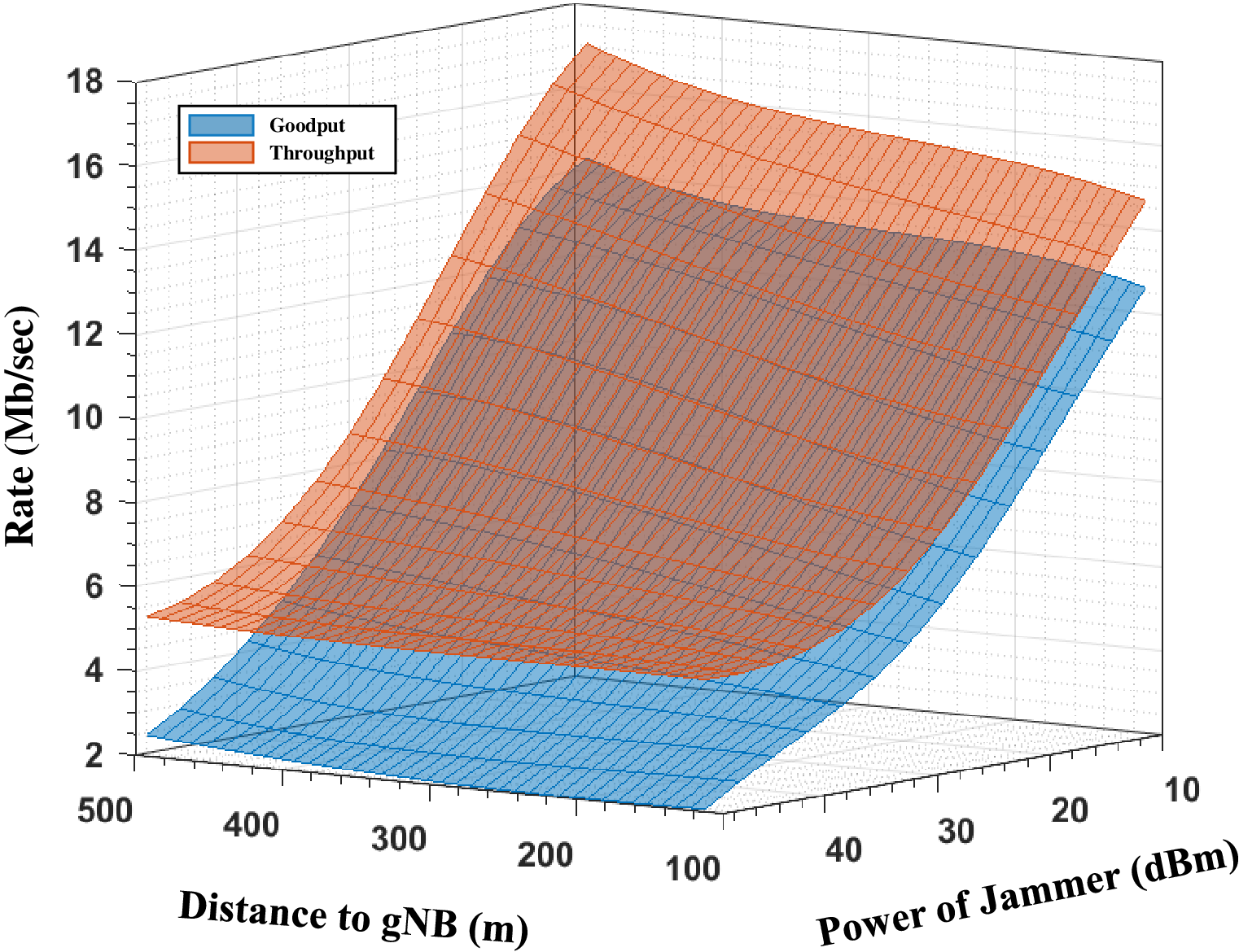}
    \caption{Cell throughput and goodput versus transmit power of the jammer and jammer-gNB distance.\vspace{-0.2in}}
    \label{fig:dgNB and Pj}
    \vspace{2mm}
\end{figure}

The relationship between 5G cell throughput and goodput concerning jamming power and jamming distance from the gNB is depicted in Fig. \ref{fig:dgNB and Pj}. Clearly, higher jamming power corresponds to lower throughput and goodput. As the jamming power increases from $10$ dBm to $40$ dBm, both throughput and goodput experience a decrease. However, when increasing the jamming power from $40$ dBm to $50$ dBm, the throughput remains relatively stable, while the goodput continues to decrease. This observation arises because throughput is measured based on the total number of packets, including re-transmissions, while goodput is calculated solely based on new arrivals.

The conclusion drawn is that increasing jamming power affects more UEs, leading to a higher number of retransmitted packets through the HARQ process. Consequently, the total throughput remains relatively constant, but the goodput decreases due to the reduction in the number of new packets in the network. This distinction emphasizes the impact of jamming power on both retransmissions and new packet arrivals, influencing the overall performance metrics differently.\vspace{-1mm}

\graphicspath{{figs/}}

\vspace{-1mm}
\subsection{SSB Vulnerability To Jamming}

In this section, we consider two types of jamming, namely, barrage and smart. Assuming $P_{RE}^{Rx}$ as the received signal power per each resource element (RE) at the UE location, the signal-to-jamming-noise ratio (SJNR) in the presence of a jammer is expressed as \vspace{-2mm}
\begin{equation}\label{Eq.SJNR}
	\begin{split}
		SJNR=\frac{P_{RE}^{Rx}}{P_j^{RE}+P_N^{RE}},
	\end{split}
\end{equation}
\noindent where $P_j^{RE}$ and $P_N^{RE}$ are the jamming and noise powers per RE at the UE side, respectively. To disrupt the link, a barrage jammer needs to inject a minimum power of $P_{j,min}^{RE}$ at the UE side per each RE to bring the SJNR below the threshold of $\gamma_{th}$. In such a case, the total power of the barrage jammer at the UE side should be equal to $P^b_j = 12 \times n_{RB} \times P_{j,min}^{RE}$.

On the other hand, disrupting communication can be more straightforward for a smart jammer targeting the SSB, which provides UE with essential information for cell selection \cite{cell_search}. To prevent the initialization of any connection, a smart jammer can inject a portion of the power, denoted as \vspace{-2mm}

\begin{equation}\label{Eq.jammer_min}
	\begin{split}
		P_{j,min}^{RE}=\frac{P^{Rx}_{RE}}{\gamma_{th}}-P_N^{RE} ,
	\end{split}
\end{equation}
\noindent on SSB bursts, thereby making the jamming process more energy-efficient\footnote{Note that a smart jammer needs to be synchronized with the gNB through a limited over-the-air processing.}. In this case, the jammer searches for PSS and SSS to locate the position of SSBs and transmits the jamming signal only over the bandwidth of $n_{RB}^{ssb} \times 12 \times \Delta f$.

Furthermore, a smart jammer does not need to transmit power throughout the entire frame since the SSB bursts in 5G are periodically transmitted for a brief time interval in every two frames. Therefore, the total power of the smart jammer at the UE side is given by $P^s_j = 12 \times n_{RB}^{ssb} \times P_{j,min}^{RE}$. It is noteworthy that if one of the SSB signals (PSS or SSS) is not successfully extracted, the UE cannot establish any connection with the gNB \cite{3GPPTS138211}.

A smart jammer can act more intelligently by specifically targeting the PBCH data or PBCH DM-RS signal. By extracting the PCI through decoding PSS and SSS, the jammer becomes aware of the location of DM-RS and PBCH. Hence, instead of targeting PSS and SSS, the jammer can transmit power at PBCH, focusing only on DM-RS and PBCH data, which is easier to disrupt than PSS and SSS since PCI extraction relies on auto-correlation, making it more robust to noise, interference, and jamming than decoding PBCH. From another perspective, under the same transmit power, a wider area is affected by the jammer.

\begin{figure}[h]
   \centering
    \includegraphics[width=.95\linewidth]{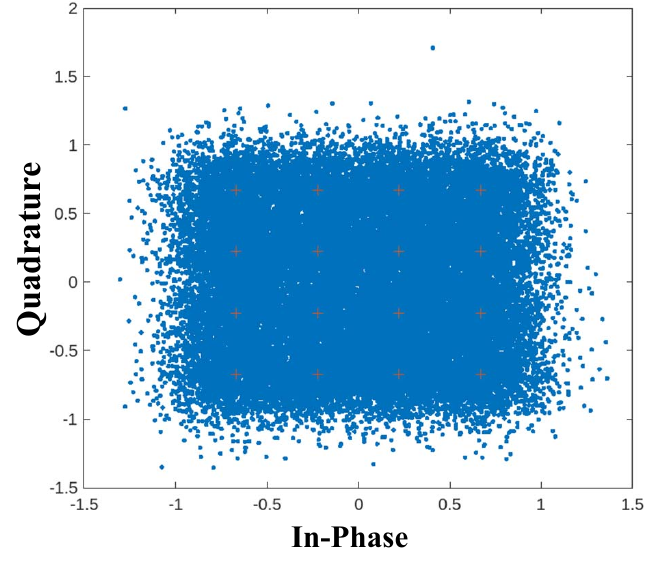 }
    \caption{Equalized PDSCH constellation - $P_j=30\,dBm$ }
    \label{fig.fig_PDSCH_cons}
\end{figure}

To validate the aforementioned discussions and examine the vulnerability of PDSCH and SSB under a jamming attack, another simulation has been configured involving both barrage and smart jammers. Assuming the gNB is located at the origin, the jammer and UE are positioned on the x-y plane at ($100,100$) and ($60,60$), respectively. Figure \ref{fig.fig_PDSCH_cons} illustrates the equalized constellation of UE's downlink data modulated with 16-QAM, with a jamming transmit power of $P_j=30$ dBm covering the entire RG. The root mean square (RMS) of error vector magnitude (EVM) is reported to be as high as $68.17\%$, as observed in the Fig. \ref{fig.fig_PDSCH_cons}.

\begin{figure}[h]
   \centering
    \includegraphics[width=.95\linewidth]{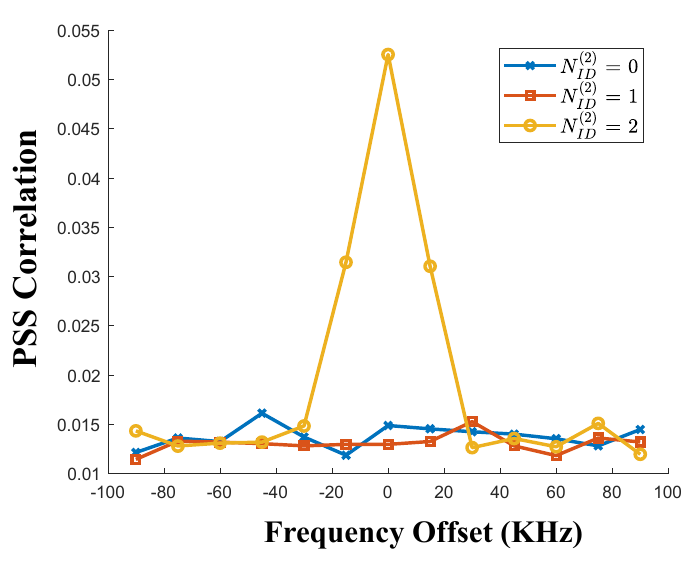 }
    \caption{PSS correlation with whole RG under jamming attack- $P^t_j=30\,dBm$}
    \label{fig.fig_PSS_1}
\end{figure}

\begin{figure}[h]
   \centering
    \includegraphics[width=.95\linewidth]{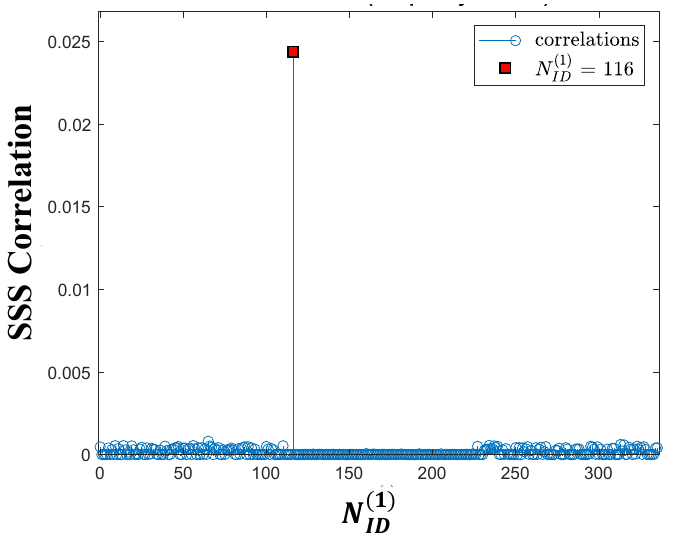 }
    \caption{PSS correlation with whole RG under jamming attack- $P^t_j=30\,dBm$}
    \label{fig.fig_SSS_1}
\end{figure}

Figs. \ref{fig.fig_PSS_1} and \ref{fig.fig_SSS_1} depict the PSS and SSS correlation under the jamming power $P_j=30\,dBm$. As it is shown, the peaks related to SSS cell ID ($N_{ID}^{(1)}$) and PSS cell ID ($N_{ID}^{(2)}$) are clear and the UE can successfully extract the PCI which has been set to 350. In the sequel, a smart jammer transmits its power only to the RBs corresponding to SSB. 
\graphicspath{{figs/}}

\graphicspath{{figs/}}

\begin{figure}[h]
   \centering
    \includegraphics[width=.95\linewidth]{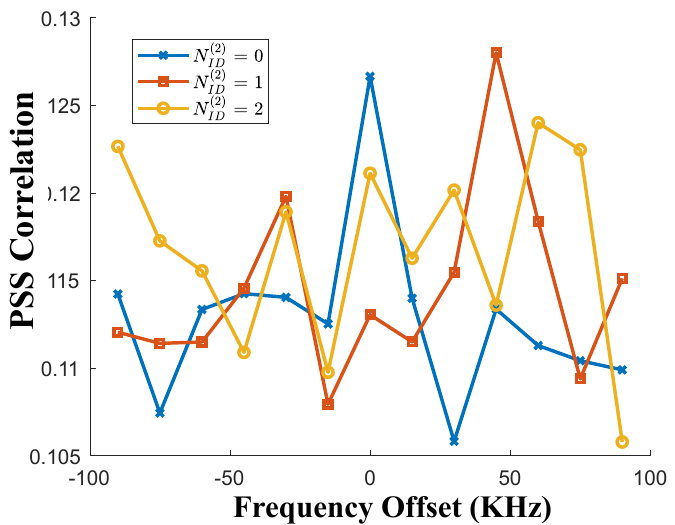 }
    \caption{PSS correlation, only SSB is under jamming attack- $P^t_j=27.5\,dBm$}
    \label{fig.fig_PSS_2}
\end{figure}

\begin{figure}[h]
   \centering
    \includegraphics[width=.95\linewidth]{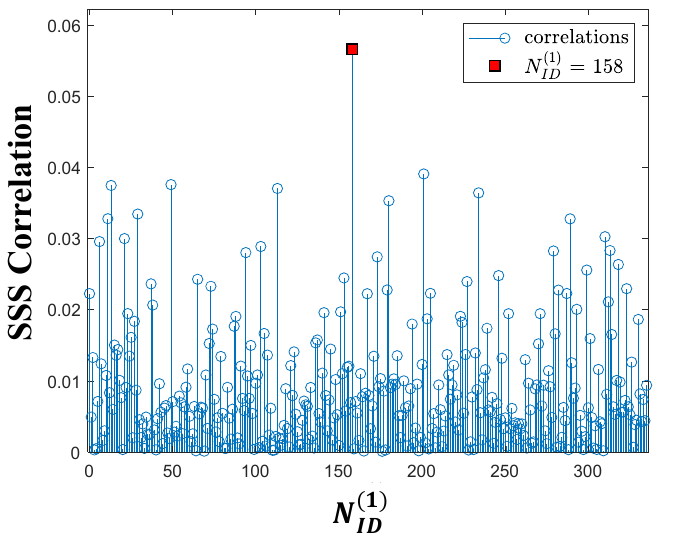 }
    \caption{SSS correlation, only SSB is under jamming attack- $P^t_j=27.5\,dBm$}
    \label{fig.fig_SSS_2}
\end{figure}

Figs. \ref{fig.fig_PSS_2} and \ref{fig.fig_SSS_2} illustrate the correlation of PSS and SSS under a smart jammer attack with transmit power equals $P_j=27.5$ dBm. The peak related to PSS is not extracted anymore in this power which is 2.5dBm lower than the barrage jammer case. Therefore, the smart jammer targeting SSB can disrupt the network with lower power, making the jamming process more energy-efficient, as discussed earlier. Note that since $n_{RB}^{ssb}/n_{RB}=0.39$ (see Table \ref{table:parameters}), the reduction in jamming power is expected to be $10\log(0.39)=-3.97$ dB, proportional to the bandwidth. However, PSS and SSS symbols are more robust to jamming compared to the PDSCH symbols\footnote{The EVM reported for PSS symbols is $57\%$, which is even less than that for PDSCH.}.

\graphicspath{{figs/}}

\graphicspath{{figs/}}

\begin{figure}[h]
   \centering
    \includegraphics[width=.95\linewidth]{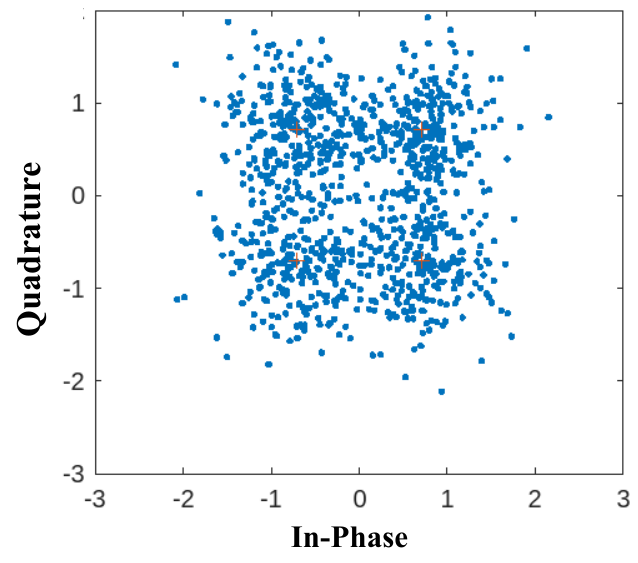 }
    \caption{Equalized PBCH constellation- $P^t_j=25\,dBm$}
    \label{fig.fig_PBCH_cons}
\end{figure}

The constellation of PBCH data is plotted in Figs. \ref{fig.fig_PBCH_cons} under jamming power of $P_j=25$ dBm in which the PCI of 350 has been extracted. However, the RMS EVM for PBCH data is reported as $51.59\%$ which is so high that the PBCH data cannot be decoded under such an attack. Since extracting PCI is based on correlation, it is more robust to impairments than PBCH data. Hence, with a jamming power of $25$ dBm, the PCI is successfully decoded, but not the PBCH data. Thus, a smart jammer can reduce its power by targeting only the PBCH data.

\graphicspath{{figs/}}
 
\begin{figure}[h]
   \centering
    \includegraphics[width=.95\linewidth]{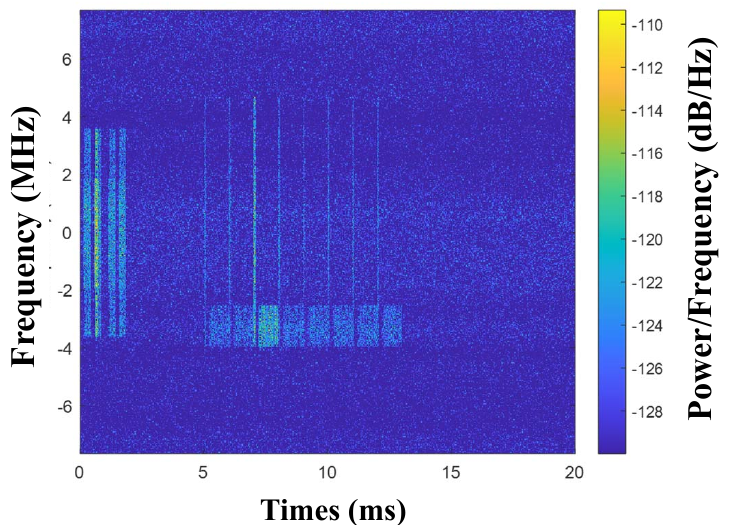 }
    \caption{Spectrogram of RG, SSB index $\#2$ boosted due to beam sweeping- $P_j=25\,dBm$}
    \label{fig.fig_Spec}
\end{figure}

The impact of beam sweeping on the smart jamming attack is considered in the following analysis. Eight SSB bursts are configured during two frames (20 ms), and the received spectrogram is plotted in Fig. \ref{fig.fig_Spec}. As can be seen, the third SSB has higher power (brighter in color) which is the effect of beam sweeping.

\begin{figure}[h]
   \centering
    \includegraphics[width=.95\linewidth]{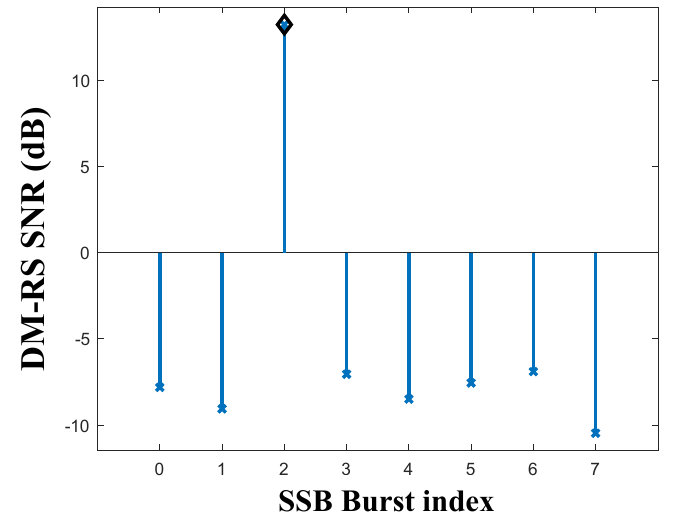 }
    \caption{DM-RS SJNR vs SSB bursts indices- $P^t_j=25\,dBm$}
    \label{fig.fig_SSB_index}
\end{figure}

\begin{figure}[h]
   \centering
    \includegraphics[width=.95\linewidth]{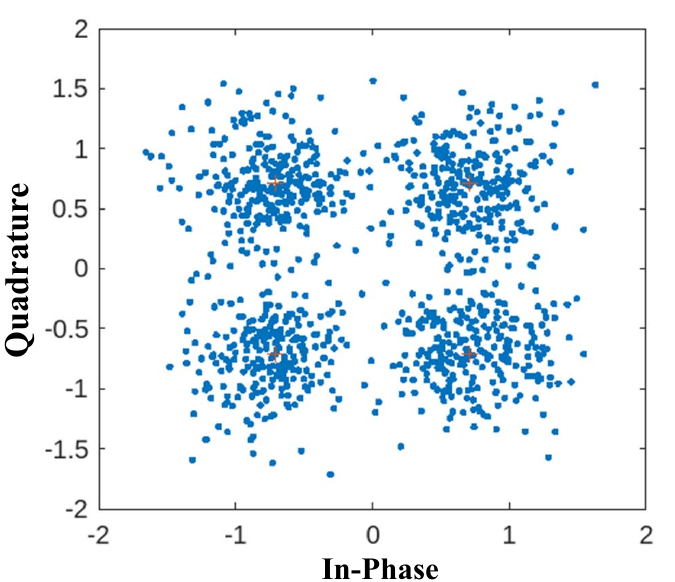 }
    \caption{Equalized PBCH constellation- $P^t_j=25\,dBm$}
    \label{fig.fig_PBCH_cons_boost}
\end{figure}

\graphicspath{{figs/}}
The estimated received SJNR for DM-RS in PBCH is plotted in Fig. \ref{fig.fig_SSB_index} versus the SSB index which shows the third SSB burst is received with higher power. The equalized PBCH data constellation is depicted in Fig. \ref{fig.fig_PBCH_cons_boost} under $P_j=25$ dBm which shows beam sweeping can make the SSB burst more robust to the SSB jamming attack with the PBCH RMS EVM is reported as $23.36\%$.

\graphicspath{{figs/}}

\graphicspath{{figs/}}

This analysis demonstrated that the information contained in SSB (including PSS correlation, SSS correlation, and PBCH data) can be effectively utilized for detecting both smart and barrage jammers. Given the varying sensitivity of these features to jamming power, fusing them into a comprehensive detection model is anticipated to significantly enhance the accuracy of jamming detection in the radio frequency (RF) domain.

 \section{Conclusion}
 \label{sec_conclusion}
This work explores the susceptibility of 5G cells to both barrage and smart jamming attacks, particularly in scenarios where user equipments (UEs), such as connected vehicles deployed for a mission, follow the spatio-temporal parametric stepping (STEPS) mobility model. The assessment of network metrics, considering jamming power, location, and varying numbers of jammers, has been thoroughly examined. Furthermore, the impact of barrage and smart jamming attacks on different aspects of the synchronization signal block (SSB) has been investigated. The numerical results demonstrate that a smart jammer, particularly one targeting SSB, proves to be more efficient than a barrage jammer or a smart jammer aimed at the primary synchronization signal (PSS) and secondary synchronization signal (SSS). Finally, the study highlights the efficiency of beam sweeping in enhancing the robustness of the cell selection process against such jamming threats.  

\section*{Acknowledgement}
\vspace{-0.08in}
This work was supported in part by funding from the Innovation for Defence Excellence and Security (IDEaS) program from the Department of National Defence (DND) and in part by the NSERC CREATE TRAVERSAL program. 


\bibliographystyle{IEEEtran}

\end{document}